\documentclass[manuscript]{aastex}
\pdfoutput=1

\addtocounter{footnote}{1}

\slugcomment{Submitted to The Astrophysical Journal}

\shorttitle{Energetic Particle Pressure at Interplanetary Shocks: STEREO-A Observations}
\shortauthors{Lario et al.}

\begin{document}


\title{Energetic Particle Pressure at Interplanetary Shocks: STEREO-A Observations}


\author{D. Lario\altaffilmark{1},  R.~B. Decker\altaffilmark{1} and E.C. Roelof\altaffilmark{1} }
\affil{The Johns Hopkins University, Applied Physics Laboratory,
    Laurel, MD 20723}

\and

\author{A.-F. Vi${\tilde{\hbox{n}}}$as\altaffilmark{2}}
\affil{NASA Goddard Space Flight Center, Greenbelt, MD 20770}



\altaffiltext{1}{The Johns Hopkins University, Applied Physics Laboratory,
    Laurel, MD 20723}
\altaffiltext{2}{NASA Goddard Space Flight Center, Greenbelt, MD 20770}


\begin{abstract}
We study periods of elevated energetic particle intensities observed by STEREO-A when the partial pressure exerted by energetic ($\geq$83 keV) protons ($P_{EP}$) is larger than the pressure exerted by the interplanetary magnetic field ($P_{B}$). In the majority of cases, these periods are associated with the passage of interplanetary shocks. Periods when $P_{EP}$ exceeds  $P_{B}$ by more than one order of magnitude are observed in the upstream region of fast interplanetary shocks where depressed magnetic field regions coincide with increases of the energetic particle intensities. When solar wind parameters are available, $P_{EP}$ also exceeds the pressure exerted by the solar wind thermal population ($P_{TH}$). Prolonged periods ($>$12 h) with both $P_{EP}$$>$$P_{B}$ and $P_{EP}$$>$$P_{TH}$ may also occur when { energetic particles accelerated by an approaching shock encounter} a region well-upstream of the shock characterized by low magnetic field magnitude and tenuous solar wind density. Quasi-exponential increases of the sum $P_{SUM}$=$P_{B}$+$P_{TH}$+$P_{EP}$ are observed in the immediate upstream region of the shocks regardless of individual changes in $P_{EP}$, $P_{B}$ and $P_{TH}$, indicating a coupling between $P_{EP}$ and the pressure of the background medium characterized by $P_{B}$ and $P_{TH}$. The quasi-exponential increase of $P_{SUM}$ implies a convected exponential radial gradient $\partial{P_{SUM}}/\partial{r}$$>$0 that results in an outward force applied to the plasma upstream of the shock. This force can be maintained by the mobile energetic particles streaming upstream of the shocks that, in the most intense events, drive electric currents able to generate diamagnetic cavities and depressed solar wind density regions.
\end{abstract}





\section{INTRODUCTION}
Contributions to the total plasma pressure measured in the interplanetary medium include: 
(i) the pressure exerted by the solar wind thermal particles, usually computed as 
$P_{TH}$=$n_{p} k T_{p}+ n_{e} k T_{e}$ 
(where $k$ is the Boltzmann constant, $n_{p}$ the solar wind proton density, $n_{e}$ is the solar wind electron density, 
$T_{p}$ the solar wind proton temperature, $T_{e}$  the solar wind electron temperature, 
and where the contribution from alpha particles and heavier ions is assumed negligible);
(ii) the hydro-static magnetic field pressure computed as $P_{B}$=$B^{2}/2\mu_{0}$ (where $B$ is the magnetic field intensity and $\mu_{0}$ is the magnetic permeability);
(iii) the pressure exerted by energetic particles $P_{EP}$ (to be defined below); and
(iv) the dominant portion of the plasma stress tensor $\rho \vec{V}\vec{V}$ 
(where $\rho$ is the mass density and $\vec{V}$ the flow velocity measured in the appropriate reference frame) 
that produces the so-called ``ram" pressure. 
This paper concerns itself only with the contribution that solar energetic particles (SEPs) make to the 
net pressure $P_{SUM}$=$P_{TH}$+$P_{B}$+$P_{EP}$ and how, 
{ within that sum}, the energetic particle pressure  $P_{EP}$ compares with $P_{B}$ and $P_{TH}$.
 
Magnetohydrodynamic studies of solar wind structures 
  generally assume that the pressure exerted by energetic particles is negligible compared to 
  $P_{TH}$ or $P_{B}$. 
  However, there have been a number of distinct periods and/or regions of interplanetary space when 
  the partial pressure exerted by energetic particles is comparable to or even exceeds both $P_{B}$ and $P_{TH}$. 
  For example, \citet{marhavilas11} used energetic particle and magnetic field measurements made by the Ulysses spacecraft from October 1990 to June 2009 
  (with heliocentric radial distances $R$ ranging from 1 to 5.4 AU and heliographic latitudes ranging from $\sim$80$^{\circ}$ south to $\sim$80$^{\circ}$ north) 
  to compare $P_{B}$ with the partial pressure $P_{EP}$ exerted by energetic protons in the energy range 20 keV-5 MeV. 
  These authors found that the periods of dominant magnetic energy ($P_{B}$$>$$P_{EP}$) prevail most of the time, but for some periods
  $P_{EP}$ was observed to exceed $P_{B}$ in some cases by more than one order of magnitude. 
  Such periods were associated with energetic particle intensity enhancements associated with the passage of shocks 
  { driven by coronal mass ejections (CMEs) or corotating interaction regions (CIRs)} 
 as well as with intervals of prominent magnetic depressions \citep[cf.][]{kasotakis99, marhavilas12}.  
  In this paper we study how common the periods with $P_{EP}$ exceeding $P_{B}$ are at heliocentric distances $R$$\sim$1 AU.

The highest SEP intensities in a solar cycle tend to be observed in association with the passage of interplanetary shocks 
\citep[e.g.,][]{reames99, lario08}. 
 Therefore,  regions of interplanetary space where $P_{EP}$ is likely to be comparable to or even exceed $P_{B}$ or $P_{TH}$
 are more likely to be found around shocks.
 { 
  A possible coupling between the pressure exerted by the energetic particles and the 
     pressure of the background medium characterized by $P_{B}$ and $P_{TH}$ could produce dynamical effects 
     on the background plasma and magnetic field \citep[e.g.,][]{drury81, axford82}. 
     For example, if there is a positive particle pressure gradient (in the direction of the flow), 
     as expected in front of shocks accompanied by intense energetic particle enhancements, 
     a slowing down of the plasma flow has been  predicted   \citep[e.g.,][]{axford77, axford82, drury81}. 
    Therefore, when the contribution of $P_{EP}$ to the total pressure is larger than the contributions of $P_{B}$ and $P_{TH}$,
    the dynamic effects produced by the energetic particles on the background medium where a shock propagates 
    should be reflected in changes of the upstream plasma and magnetic field parameters.
    The variation of the plasma conditions upstream of a shock  leads also to a modification of the 
    jump conditions across the shock and hence a variation of the  properties of the shock relative to those of a shock propagating into a medium dominated by 
 the pressures of  the background magnetic field and thermal plasma \citep[e.g.,][]{blandford80}.}


{ In this paper we select periods where $P_{EP}$ exceeds $P_{B}$,  we analyze whether
 the effects of a dominant $P_{EP}$ are reflected on the evolution of the plasma and magnetic field parameters, and 
we determine whether there is a relationship between the evolution of $P_{B}$, $P_{TH}$ and $P_{EP}$. }
  We should indicate that, since this is an observation-based study, energetic particle instruments can measure only a partial energetic particle pressure 
  $P_{EP}$ because their energy range is limited at both the low and high ends. 
  However, it is obvious that the quantity we indicate by $P_{EP}$ 
  (i.e. the partial pressure exerted by energetic particles) provides a rigorous lower bound on 
  the total pressure exerted by the whole energetic particle population.
We should also indicate that strictly speaking, all such pressures  are, by definition, properly calculated in the plasma frames (upstream or downstream of the shocks) and,
  when evaluating MHD conservation equations at a shock, 
  the ``ram" pressure is computed in the appropriate shock frame. 
  { The correct computation of the ram pressure requires an accurate determination of the shock parameters 
  (among them the direction of the normal to the shock and the shock speed). 
  For those shocks where 
  $P_{EP}$ is comparable to or exceeds $P_{B}$ and $P_{TH}$, the  contribution of 
   $P_{EP}$ may play a significant role 
   in the computation of the shock parameters, especially in the conservation of momentum and energy flux across the shock  
\citep[e.g.,][]{lario15}.
Current methods commonly applied to compute shock parameters do not consider the contribution of $P_{EP}$ into the Rankine-Hugoniot conservation conditions \citep[e.g.,][]{vinas86, szabo94}.
Therefore, for the shocks analyzed in this article, 
 the evolution of the  ``ram" pressure will not be included in our analyses.
 Whereas the ``ram" pressure usually dominates the total pressure measured upstream of shocks \citep[e.g.,][]{lario15},
the presence of a significant energetic particle pressure $P_{EP}$, together with a 
 possible coupling between $P_{EP}$,   $P_{TH}$ and $P_{B}$, may modify the upstream region of the shocks
 and thus affect the jump conditions across the shocks.}
  In this paper we { focus the study on} the evolution of the sum of pressures $P_{SUM}$=$P_{B}$+$P_{TH}$+$P_{EP}$ 
  in the vicinity of shocks for those periods where the partial pressure of energetic particles exceeds $P_{B}$.

\section{TWO LARGE SHOCK EVENTS AT $\sim$1 AU WITH $P_{EP}$$>$$P_{B}$}

Significant cases where $P_{EP}$ exceeded $P_{B}$ by more than one order of magnitude at $\sim$1~AU include
 the events on 20 October 1989 observed near Earth \citep{lario02} and 
 on 23 July 2012 observed by the
 Spacecraft-A of the Solar Terrestrial Relations Observatory
(henceforth STEREO-A) { located} at $R$=0.96 AU { and 121$^{\circ}$ west of Earth} \citep{russell13}.
{ Both events} were associated with the passage of interplanetary shocks. 
 Figure~1 shows from top to bottom (a) energetic proton intensities measured by 
 (left) the Energetic Particle Sensor (EPS) on the Geostationary Operational Environmental Satelite  \citep[GOES-7/EPS;][]{sauer93} on day 293 of 1989 (i.e. 20 October 1989) and 
 (right) by the High Energy Telescope (HET) of the In-situ Measurement of Particles and CME Transients (IMPACT) suite of instruments
 on board STEREO-A \citep{rosenvinge08} on day 205 of 2012 (i.e. 23 July 2012); 
 (b) magnetic field magnitude measured by (left) the Goddard Space Flight Center (GSFC) magnetometer on board the Interplanetary Monitoring Platform  \citep[IMP-8;][]{scearce76} and
  (right) the Magnetic Field Experiment of the IMPACT instrument suite \citep{acuna08} on board STEREO-A; 
  (c) solar wind proton density $n_{p}$, (d) solar wind proton temperature $T_{p}$, and 
  (e) solar wind speed $V_{sw}$ as measured by (left) the Massachusetts Institute of Technology (MIT) Faraday Cup on board IMP-8 \citep{scearce76}
    and (right) by the PLASTIC sensor on board STEREO-A \citep{galvin08}. 
    Figure~1f shows $P_{B}$ (red symbols) and the solar wind thermal pressure (blue symbols) 
    computed as $P_{TH}$=$n_{p} k (T_{p}+T_{e})$ \citep[where we have assumed charge neutrality $n_{p}$=$n_{e}$,
    negligible contribution of heavy ions, 
    and $T_{e}$=2 $T_{p}$ based on statistical surveys of proton and electron temperatures in post-shock plasmas, cf.][]{gosling87}. 
    The black trace in Figure 1f shows the energetic particle partial pressure $P_{EP}$ computed as
    $P_{EP}$=$(4\pi/3)(2m)^{1/2}\int_{E_{1}}^{E_{2}} dE\, E^{1/2} j(E)$  , where $E$ is the proton kinetic energy, 
    $j(E)$ is the proton differential flux, $m$ is the proton mass, and $E_{1}$ and $E_{2}$ are the limits of the instrumental energy 
    range over which $P_{EP}$ is computed (as indicated in Figure 1f). The purple traces in Figure 1f show the sum $P_{B}$+$P_{TH}$+$P_{EP}$=$P_{SUM}$.
    
The solid vertical lines in Figure~1 (labeled with the symbol S) indicate the passage of interplanetary shocks with which these intense particle events were associated \citep{cane95, russell13}.
    The vertical dotted lines in Figure 1 (labeled with the number 2) indicate the onset of an abrupt depression of the magnetic field 
    coincident with an increase in the energetic proton intensities observed before the passage of the shocks. 
    The dashed vertical lines in Figure 1 (labeled with the number 1) indicate a small increase of magnetic field magnitude, 
    solar wind density and temperature that could be associated with an earlier weak shock \citep[cf.][]{cane95, lario02, russell13}.
    
 Similarities between the two events shown in Figure 1 have been discussed by \citet{freed14}. 
    Here we point out that in both events: 
    (1) the increase in the energetic particle intensity coincided with a reduction of the magnetic field intensity well before the arrival of the shocks; 
    (2) the period before the arrival of the shocks was characterized by a steady increase of the solar wind speed $V_{sw}$; 
    (3) just before the arrival of the shocks and for a period of $\sim$30 minutes magnetic field strength started to increase; and 
    (4) the energetic particle pressure $P_{EP}$ exceeded the magnetic field pressure $P_{B}$ 
    by more than one order of magnitude during an extended period of $\sim$195 minutes and $\sim$96 minutes 
    before the passage of the shock on 20 October 1989 (day of year 293) and 23 July 2012 (day of year 205), respectively. 
    The availability of solar wind data throughout the event on 20 October 1989 allows us to see that: 
    (i) during the period of depressed magnetic field the energetic particle pressure also exceeded the thermal pressure $P_{TH}$; 
    (ii) the period of $\sim$30 minutes before the shock where $B$ increases shows also an increase of $n_{p}$; 
    (iii) $P_{EP}$ { continued to increase for the whole period where  $P_{B}$ and $P_{TH}$ were depressed}; and 
    (iv) the sum of the pressures $P_{B}$+$P_{TH}$+$P_{EP}$ showed a gradual increase as the shock approached the observer \citep{lario02}. 
    Unfortunately, the lack of solar wind data from STEREO-A during this intense particle event does not allow us to compute $P_{TH}$ or $P_{SUM}$ 
    during the passage of the shock on 23 July 2012. 
    We would { like to point out} that the
     two events shown in Figure 1 were associated with intense solar eruptions 
    able to generate fast solar wind disturbances traveling from the Sun to $\sim$1 AU with very short transit times 
    ($\sim$28 hours for the shock on 20 October 1989 and $\sim$18 hours for the shock on 23 July 2012; 
 corresponding to average transit speeds of $\sim$1490 km s$^{-1}$ and $\sim$2225 km s$^{-1}$, respectively). 
    Therefore, these two events were associated with very fast shocks \citep{freed14}.
{ As described in the introduction, signatures of the possible coupling between $P_{EP}$ and the 
     pressure of the background medium characterized by $P_{B}$ and $P_{TH}$
     include the slowing down of the plasma flow in front of the shocks
\citep[e.g.,][]{axford77, axford82, drury81}.     }     
     In the two events shown in Figure 1, the flow velocity $V_{sw}$ in the spacecraft frame shows a gradual increase
      before the shock passage (Figure~1e) that translates into a decrease of the flow speed in the shock rest frame 
      \citep[see also][]{terasawa99}.

{ In the following section we select periods observed by STEREO-A over the first half of solar cycle 24 where the
pressure exerted by energetic particles exceeds $P_{B}$}, and determine whether the expected effects of $P_{EP}$
on the plasma and magnetic field parameters are actually observed, and whether there is a
relationship between the evolution of $P_{B}$, $P_{TH}$, and $P_{EP}$ .

\section{STEREO-A OBSERVATIONS}

We have examined STEREO-A data from the beginning of 2009 to day 186 of 2014 to study
 how frequent are the periods with both $P_{EP}$$>$$P_{B}$ and $P_{EP}$$>$$P_{TH}$. 
 The time interval examined covers both the rising and maximum phases of solar cycle 24. 
 As survey data for selecting our events we use: hourly averages of the magnetic field magnitude $B$ measured by the magnetometer of the IMPACT 
 instrument suite \citep{luhmann08} and downloaded from www-ssc.igpp.ucla.edu/ssc/stereo/; 
 the solar wind proton density $n_{p}$ and temperature $T_{p}$ as measured by the PLASTIC instrument \citep{galvin08} 
 and downloaded from stereo.sr.unh.edu; and the energetic ion intensities measured by the 
 Solar Electron and Proton Telescope (SEPT) of the IMPACT instrument suite \citep{muller08} and downloaded from www2.physik.uni-kiel.de/stereo/. 
 The hourly averages of the logarithms of the differential particle intensities $j(E)$ measured by SEPT have been fitted 
 { using a functional form described as a power-law with an exponential cutoff}
  over the energy range 83 keV-2.22 MeV, 
 { and considering} only those points when the differential intensity of the energy channel 1.98-2.22 MeV was above 0.05 particles (cm$^{2}$ s sr MeV)$^{-1}$. { This intensity threshold  allows us to consider only cases where intensities are above instrumental background and the energy spectra 
 over the differential energy channels of STEREO-A/IMPACT/SEPT are well fitted by the chosen functional form}.
  The energetic particle partial pressure $P_{EP}$ was calculated over the energy interval $E_{1}$=83 keV and $E_{2}$=2 MeV, 
  assuming that the STEREO-A/IMPACT/SEPT detector responded only to protons. 
  This assumption gives us a lower limit for $P_{EP}$, especially when heavier ions make an important contribution to the measured particle fluxes. 
  The limited energy interval 83-2000 keV also gives a lower limit of the actual pressure exerted by a complete population of energetic particles that, 
  in principle, may extend over a broader energy range, e.g., down to pickup ion energies ($^{>}_{\sim}$keV)
  and well above 2 MeV (especially during intense SEP events).
  
Figure 2 shows from top to bottom hourly averages of: 
  (a) 1.98-2.22 MeV ion differential intensities measured by STEREO-A/IMPACT/SEPT 
  (the horizontal solid line at 0.05 particles (cm$^{2}$ s sr MeV)$^{-1}$ indicates the threshold value above which $P_{EP}$ has been computed); 
  (b) $P_{EP}$ computed over the energy range 83-2000 keV; (c) $P_{B}$=$B^{2}/2\mu_{0}$; (d) $P_{TH}$=$n_{p} k(T_{p}+T_{e})$ 
  (computed assuming $T_{e}$=2 $T_{p}$); (e) the ratio $P_{EP}$/$P_{TH}$; and (f) the ratio $P_{EP}$/$P_{B}$. 
  The horizontal solid lines in Figures 2e and 2f indicate the values $P_{EP}$/$P_{TH}$=1 and $P_{EP}$/$P_{B}$=1, 
  respectively. Table~1 lists the time intervals when $P_{EP}$/$P_{B}$ $>$1. 
  The first column of Table 1 lists the year and day of year when the periods with $P_{EP}$$>$$P_{B}$ were observed. 
  As expected, all these periods are characterized by elevated energetic particle intensities due to the occurrence of intense solar energetic particle (SEP) events. 
  The second column lists the duration of the time intervals when $P_{EP}$$>$$P_{B}$ (identified using one-minute averaged data as described below in the following sections). 
  The third column of Table 1 indicates whether the periods with $P_{EP}$$>$$P_{B}$ were associated with the passage of an interplanetary shock. 
  With the exception of one case (2012/025) where the local passage of an interplanetary shock could not be identified, all periods with $P_{EP}$$>$$P_{B}$ 
  occurred in the upstream region of interplanetary shocks. For those cases associated with interplanetary shocks, 
  the fourth column of Table 1 gives the magnetic field compression ratio $r_{B}$=$B_{d}/B_{u}$ and the density compression ratio $r_{n}$=$n_{d}/n_{u}$ 
  (where the subscripts $d$ and $u$ indicate the values of $B$ and solar wind proton density measured downstream and upstream of the shock, respectively). 
  In order to obtain $r_{B}$ and $r_{n}$ we have used the shock analysis tool developed by \citet{vinas86}. 
  Note that for the single event on 23 July 2012 (2012/205), the \citet{vinas86} method was not applied 
  because of the lack of solar wind data, and we only list the magnetic compression ratio 
  as provided by J. Lian (www-ssc.igpp.ucla.edu/$\sim$jlian/STEREO/Level3/). 
  Note that in the application of the \citet{vinas86} method
  $P_{EP}$ was not included in the momentum conservation equations and
 the energy equation of the Rankine-Hugoniot conservation relations across the shock
was not considered.
Under this assumption, density ratios $r_{n}$$>$4 can be obtained.
{ For those cases associated with shocks, and where $P_{SUM}$=$P_{EP}$+$P_{B}$+$P_{TH}$ can
be computed, we list in the last column of Table~1 the time scale of the exponential increase observed before the shock passage in the time evolution of  $P_{SUM}$ (see discussion below).}

The short-duration ($\leq$100 minutes) periods on 2012/149 and 2013/206 (see Table~1) were associated with 
extremely strong shocks with large compression ratios, whereas the shocks on 2011/156, 2012/315 and 2013/234, 
observed in association with the long-duration ($>$12 h) $P_{EP}$$>$$P_{B}$ periods, had significantly lower compression ratios. 
The period on 2012/025 occurred in a region with depressed magnetic field observed just before the passage of the interplanetary 
counterpart of a coronal mass ejection (i.e. an CME), but no interplanetary shock could be identified. 
In the following sections we analyze in detail the time intervals listed in Table~1.
We start with the period not associated with a shock that constitutes the exception among the periods with $P_{EP}$$>$$P_{B}$ analyzed in this paper.

\subsection{The period on 2012/025 associated with a magnetic field depression}

Figure 3 shows, from top to bottom, one-minute averages of 
(a) 1.4-1.6 MeV ion intensities observed by STEREO-A/IMPACT/STEP, 
(b) magnetic field magnitude, solar wind proton (c) density, (d) temperature, and (e) speed. 
Figure 3f compares $P_{B}$ (red dots), $P_{TH}$ (blue dots), the partial pressure $P_{EP}$ (black trace), 
and the sum $P_{SUM}$=$P_{B}$+$P_{TH}$+$P_{EP}$ (purple trace). 
$P_{EP}$ has been computed over the indicated energy range ($E_{1}$=83 keV, $E_{2}$=77.6 MeV) that includes all the proton differential 
energy channels of STEREO-A/IMPACT/STEP \citep{muller08}, STEREO-A/IMPACT/LET \citep{mewaldt08} and STEREO-A/IMPACT/HET \citep{rosenvinge08} 
that showed particle intensity enhancements
(the value of $E_{2}$ corresponds to the geometric mean energy of the highest proton energy channel of HET). 

The intense energetic particle population observed during this period has been associated with a SEP event 
generated by a fast halo CME 
(plane-of-sky speed $V_{CME}$=2175 km s$^{-1}$ according to the LASCO/CME catalog at cdaw.gsfc.nasa.gov/CME$\_$list) 
at 2012/023/04:00 UT \citep[e.g.,][]{park13, richardson14}. 
The vertical dashed line in Figure 3 indicates the leading edge of an ICME 
as identified by J. Lian (www-ssc.igpp.ucla.edu/$\sim$jlian/STEREO/Level3/). 
There was no detectable change in the ion intensity-time profiles upon the entry of STEREO-A into this ICME 
\citep[cf. Fig. 22 of][]{park13}. Prior to the arrival of the ICME, particle intensities were already elevated 
because of the solar eruption on day 2012/023. 
A gradual, but significant particle intensity enhancement was observed starting on day $\sim$25.35. 
After $\sim$4 hours the rise was terminated by a decrease of magnetic field magnitude  
(indicated by the vertical dotted line in Figure 3). 
The period between the dotted vertical line and the entry of the spacecraft into the ICME (dashed vertical line)
 was characterized by elevated particle intensities; it started with comparable values between $P_{EP}$ and $P_{B}$, 
 but was followed after $\sim$5 h by a prolonged period of $\sim$288 minutes with decreased magnetic field and $P_{EP}$$>$$P_{B}$. 
 The elevated solar wind densities observed throughout the time interval with $P_{EP}$$>$$P_{B}$, however, implied that $P_{TH}$$>$$P_{EP}$. 
 Throughout all this period where $P_{EP}$ reached a plateau, $P_{SUM}$ (purple symbols in Figure 3f) was dominated by $P_{TH}$ and remained approximately constant. 
 The entry of the spacecraft into the ICME (vertical dashed line in Figure~3) was marked by an increase of the magnetic field strength,
 a decrease of $P_{TH}$,
  but no change in particle intensities, resulting in $P_{B}$$>$$P_{EP}$, and with $P_{B}$ contributing the most to $P_{SUM}$. 
  Nevertheless, $P_{SUM}$ stayed practically constant throughout this event (showing similar values before and after the entry 
  of the spacecraft into the ICME) even though $P_{SUM}$ was dominated by $P_{TH}$ before and by $P_{B}$ after entering the ICME (dashed vertical line), 
  consistent with the ICME being in pressure balance with its plasma environment.

\subsection{The periods on 2012/149 and 2013/206 associated with strong interplanetary shocks}

Figure 4 shows, in the same format as Figure 3, the period associated with 
the passage of an extremely strong interplanetary shock at 2012/149/02:48 UT 
(indicated by the solid vertical line labeled by the symbol S). 
This shock was most likely associated with a fast halo CME on the Sun at 2012/147/20:57 UT 
($V_{CME}$=1966 km s$^{-1}$) that generated an intense SEP event \citep{richardson14}. 
The association between the CME at the Sun and the shock at STEREO-A (at R=0.96 AU) implies an average transit speed of $\sim$1336 km s$^{-1}$ 
between the Sun and spacecraft. 
Particle fluxes during this SEP event reached a maximum peak intensity coinciding with the arrival of the shock. 
The onset of the particle intensity increase observed just before the shock passage (indicated by the vertical dotted line in Figure 4) 
coincided with a decrease of both the magnetic field magnitude ($<$5 nT) and the solar wind density ($<$3 cm$^{-3}$) 
but with an increase in the solar wind proton temperature. 
Figure 4f shows that the period between the vertical dotted line and the arrival of the shock (of $\sim$60 minutes of duration) 
was clearly dominated by the pressure $P_{EP}$ exerted by energetic protons in the energy range 
$E_{1}$=83 keV and $E_{2}$=60 MeV. 
$P_{SUM}$ (purple symbols in Figure 4f) showed a gradual exponential increase before the arrival of the shock that started even 
before the period with $P_{EP}$$>$$P_{B}$ (we indicate this exponential increase with the dashed green line in Figure 4f). 
{ The last column of Table~1 lists the time scale $\tau$ of the exponential increase of $P_{SUM}$ obtained
from fitting the function $P_{SUM}(t)$=$P_{SUM}(t_{1})\exp{\Bigl\{(t-t_{1})/\tau\Bigr\}}$ to the time profile of $P_{SUM}$  
over the time interval indicated by the dashed green line in Figure~4f (i.e., $\sim$2.2 h before the arrival of the shock)}. 
The rise of $P_{SUM}$ was exponential even before it became dominated by $P_{EP}$ (vertical dashed line). 
The solar wind speed $V_{sw}$ also showed an increase during this period, 
but it first reached a maximum 27 minutes before the shock passage and then decreased to a local minimum observed 5 minutes before the shock. 
An increase in $V_{sw}$, $T_{p}$ and $B$ was observed for 5 minutes just before the shock. 
The time profiles of this event resemble those of the intense events on 20 October 1989 and 23 July 2012 shown in Figure 1, 
with the exceptions that: the onset of the particle increase observed before the shock arrival (at the time of the dotted vertical line) was more gradual; 
the solar wind temperature increased instead of decreasing; and the solar wind speed did not show a gradual long-lasting steady
 increase as observed in the case of the events in Figure 1.
 
Figure 5 shows, in the same format as Figure 3, the period associated with the passage of a strong shock on 2013/206/06:12 UT. 
 The elevated particle intensities observed during this event have been associated with a fast halo CME at 2013/203/06:24 UT ($V_{CME}$=1044 km s$^{-1}$) 
 \citep{richardson14}. 
 The association between the CME at the Sun and the shock at STEREO-A gives an average transit shock speed of $\sim$560 km s$^{-1}$. 
 The elevated particle intensities observed throughout this event, together with the low magnetic field magnitude ($\sim$2 nT) and low temperature ($<$10$^{5}$ K), 
 yielded energetic particle pressure exerted by 83-7586 keV protons 
 very close to both $P_{B}$ and $P_{TH}$ for more than 12 hours before the arrival of the shock. 
 The particle intensity increase observed just before the arrival of the shock coincided with a depression
  of the solar wind density and low values of the magnetic field magnitude (indicated by the vertical dotted line in Figure 5) 
  where $P_{EP}$ clearly exceeded both $P_{B}$ and $P_{TH}$. Particle intensities did not peak at the time of the shock passage
   but showed two peaks upstream and downstream of the shock. 
   The first peak yielded an increase of $P_{EP}$ and a period of $\sim$100 minutes with $P_{EP}$$>$$P_{B}$ and $P_{EP}$$>$$P_{TH}$. 
   The period dominated by $P_{EP}$ was characterized by an exponential increase of $P_{SUM}$ until the arrival of the shock
    (indicated by the dashed green line in Figure 5f). { The  e-folding time for the exponential increase of $P_{SUM}$ was $\tau$=0.097 days (Table~1)}.
    Note that although particle intensities did not peak at the time of the shock passage, the increase of $n_{p}$ and $B$ observed for $\sim$30 minutes 
    before the shock translated into the continuous exponential increase of $P_{SUM}$. 
    This increase of $n_{p}$ and $B$ just a few minutes before the shock passage is similar to that observed in the event on 20 October 1989 (left panels of Figure 1) 
    and results in lower values of $r_{B}$ and $r_{n}$ than the compression ratios that would have been obtained
     if the upstream region had been characterized by the minimum values of $B$ and $n_{p}$ observed several minutes before the shocks.

\subsection{Extended periods with $P_{EP}$ $>$ $P_{B}$ observed before the arrival of interplanetary shocks}

Figure 6 shows, in the same format as Figure 3, the time intervals associated 
with the passage of the shocks observed by STEREO-A, from left to right, at 2011/156/18:59 UT, 2012/315/22:30 UT, and 2013/234/07:05 UT. 
The solid vertical lines labeled with the symbol S indicate the passages of the shocks by STEREO-A. 
These three shocks were observed in association with intense SEP events whose particle intensity-time 
profiles maximized around the arrival of the shocks. 
The time profiles of $P_{EP}$ during these events continuously increased as the shocks approached the spacecraft. 
The periods with $P_{EP}$$>$$P_{B}$ were very extended in time, with prolonged durations of $\sim$12.5 h, 14.7 h and 22.9 h, 
for the events on days 2011/156, 2012/315 and 2013/234, respectively. 

The intense SEP event observed by STEREO-A in association with the passage 
of the shock at 2011/156/18:59 UT (left panels in Figure 6) has been analyzed in detail by \citet{lario13a, lario13b}. 
Its solar origin was an extremely fast halo CME ($V_{CME}$=2425 km s$^{-1}$) seen by LASCO at 2011/155/22:05 UT (cdaw.gsfc.nasa.gov/CME$\_$list/). 
The shock observed by STEREO-A was most likely related to the merging of this CME with a prior halo 
CME on the Sun at 2011/155/06:48 UT ($V_{CME}$=1407 km s$^{-1}$) \citep{lario13b}. 
The SEP event at STEREO-A occurred in a period of low magnetic field magnitude ($^{<}_{\sim}$5 nT) \citep[cf. Figure 7b in][]{lario13a} 
that favored the domination of $P_{EP}$ over $P_{B}$. 
Additionally, the low solar wind density ($<$3 cm$^{-3}$) observed prior to the arrival of the shock also yielded a prolonged period with $P_{EP}$$>$$P_{TH}$. 
$P_{SUM}$ also showed an exponential increase for a period of $\sim$3 h before the arrival of the shock (indicated by the dashed green line
{ with an e-folding time $\tau$=0.318 days as listed in Table~1}).

The SEP event observed by STEREO-A in association with the passage of the shock at 2012/315/22:30 UT (center panels in Figure 6) 
has been associated with a halo CME seen by LASCO at 2012/313/11:00 UT ($V_{CME}$=972 km s$^{-1}$) \citep{richardson14}, 
implying an average transit shock speed to travel from the Sun to STEREO-A of 676 km s$^{-1}$. 
The low magnetic field magnitude ($<$5 nT), low solar wind density ($<$4 cm$^{-3}$) and low solar wind proton temperature ($<$10$^{5}$ K) 
observed before the shock arrival, together with the { presence of an intense population of shock-accelerated particles}, 
led to an extended period ($\sim$14.7 h) with $P_{EP}$$>$$P_{B}$ and $P_{EP}$$>$$P_{TH}$. 
Again, $P_{SUM}$ was observed to continuously increase throughout the upstream region of the shock with a 
final $\sim$1.6 h exponential increase just before the shock (indicated by the dashed green line { and e-folding time $\tau$=0.088 days}), 
primarily because $P_{SUM}$ was dominated by the exponential increase in $P_{EP}$.

The SEP event observed by STEREO-A in association with the passage of the shock at 2013/234/07:05 UT (right panels in Figure 6) 
has been associated with a halo CME seen by LASCO at 2013/231/23:12 UT ($V_{CME}$=877 km s$^{-1}$) \citep{richardson14}, 
implying an average transit shock speed to travel from the Sun to STEREO-A of 719 km s$^{-1}$. 
This event also occurred in a period characterized by low magnetic field magnitude ($<$5 nT) and low solar wind proton temperatures ($<$10$^{5}$ K).
The elevated particle intensities observed in the upstream region of the shock led to a period with $P_{EP}$$>$$P_{TH}$, $P_{EP}$$>$$P_{B}$ and a continuous increase of
 $P_{SUM}$ that peaked in coincidence with the arrival of the shock. Significantly, this smooth increase
of   $P_{SUM}$ was a consequence of the time profiles of both the solar wind density and the magnetic field magnitude
   showing oscillations that mirrored those of the energetic particle intensity-time profile. 
   With the exception of the last increase observed just before the arrival of the shock, abrupt increases
    of energetic particle intensities coincided with  depressions of $n_{p}$ and $B$, 
    whereas depressions of energetic particle intensities coincided with increases of $n_{p}$ and $B$. 
    These simultaneous changes in energetic particle intensities and both $n_{p}$ and $B$ are similar
     to the oscillations found in intensity-time profiles of SEP events produced by magnetic 
     discontinuities that act as barriers to the propagation of SEPs \citep[e.g.,][]{sanderson00}. 
      Similarly to the previous events, $P_{SUM}$ showed an exponential increase 
      (indicated by the dashed green line { and e-folding time $\tau$=0.235 days})
       for a period of $\sim$5 h before the shock arrival, despite the quite different individual variations of $P_{EP}$, $P_{TH}$ and $P_{B}$. 
       Their cooperative contributions thus produced a radial gradient in the pressure that exponentially increased upstream of the shock.

\subsection{Short-duration intervals associated with the passage of shocks}

Finally, we would like to point out numerous short-duration periods with $P_{EP}$$>$$P_{B}$ or 
$P_{EP}$$\sim$$P_{B}$ that are clearly discernible when using one-minute averaged data but that 
did not show up exceeding the line $P_{EP}$/$P_{B}$=1 in the hourly averages shown in Figure 2f. 
These cases are associated with interplanetary shocks in which the time interval with $P_{EP}$$>$$P_{B}$ is only reached 
very close to the shock passage. 
Figure 7 shows three examples of these shocks showing short-duration ($\sim$20 minutes) spikes in the $P_{EP}$ time profiles. 
The particle intensity increases associated with the shock passages lead to  local increases of $P_{EP}$. The periods with $P_{EP}$$>$$P_{B}$ 
are observed before the passage of the shocks and their duration is usually short. 
In the examples shown in Figure 7, the periods with $P_{EP}$$>$$P_{B}$ lasted for $\sim$21, $\sim$20, and $\sim$12 minutes 
before the passage of the shocks on 2012/029/13:04 UT, 2012/139/12:43 UT and 2013/067/02:20 UT, respectively. 
The decrease of $n_{p}$ observed just before the shocks on 2012/029 and 2013/067 also led to short periods where $P_{EP}$ exceeded $P_{TH}$. 
In all three cases, the increase in $P_{SUM}$ was only observed for a few minutes just before the shock passage (indicated by the green lines in Figure 7f
{ with e-folding time scales of 0.04 days, 0.06 days,  and 0.09 days
for the events on 2012/029, 2012/139, and 2013/067, respectively}).

\section{DISCUSSION AND SUMMARY}

Energetic particle, solar wind and magnetic field measurements from STEREO-A 
during the rising and maximum phases of solar cycle 24 show a
 large number of time intervals with both $P_{EP}$$^{>}_{\sim}$$P_{TH}$ and $P_{EP}$$^{>}_{\sim}$$P_{B}$ (Figure~2).
 When using hourly averages of all data measurements 
 we found seven time intervals of durations $>$1 h  where  $P_{EP}$$>$$P_{B}$ (Table~1).
 With the exception of one event in which $P_{EP}$$>$$P_{B}$ but $P_{EP}$$<$$P_{TH}$ (2012/025; Figure 3), 
 all the periods with $P_{EP}$$>$$P_{B}$ 
 were associated with the passage of interplanetary shocks and showed also  $P_{EP}$$>$$P_{TH}$
 (with the possible exception of the event on 2012/205 (Figure 1)  when the solar wind parameters required to compute $P_{TH}$ were not available).
 
 The most intense periods at $\sim$1 AU where $P_{EP}$ exceeded $P_{B}$ by more than one order of magnitude were 
 observed in one extreme event in solar cycle 22 
 \citep[20 October 1989;][]{lario02} and another extreme event in solar cycle 24 \citep[23 July 2012;][]{russell13}
and both occurred  in the upstream region of fast interplanetary shocks (Figure 1). 
 The upstream regions of these shocks were characterized by depressed magnetic field
 (and, when solar wind data were available, also by depleted solar wind density). 
 The decrease of the magnetic field magnitude observed before the shock coincided with an abrupt increase
  in the energetic particle intensity (dotted lines in Figure 1). 
  Downstream of the shocks, the pressure was dominated by that of the shocked thermal plasma or magnetic field.
  
  For less intense events observed   in association
  with the passage of interplanetary shocks (cf. Figures~4 and~5) a similar effect is observed
  with depressions of magnetic field and solar wind density in front of the shocks. 
In these cases, however, the depletions of the solar wind density and magnetic field magnitude were not as extreme as in the events shown in Figure 1.
Yet, in the events shown in Figures 4 and 5, the partial pressure exerted by the energetic particles still could exceed $P_{B}$ and $P_{TH}$ by a factor of 5-10.  
  
 The local enhancement of energetic particle intensities associated with the passage of shocks in the form of short-duration ($\sim$20 minutes) 
       spikes also leads to the observations of very brief periods just before the shocks with $P_{EP}$$>$$P_{TH}$ and $P_{EP}$$>$$P_{B}$ (Figure 7). 
 Nonetheless, there were   weak  depletions of the solar wind density and magnetic field  (e.g. 2012/029 and 2013/067 in Figure 7 immediately before the shocks ).
 Since the enhancement in $P_{EP}$ was very local to the shock, at the time of observation,
 it may be that the local density and field depressions were varying with time.
        
 During intense SEP events, well before the arrival of the associated shocks,
                it is possible to observe long periods ($>$12 h) with $P_{EP}$$>$$P_{TH}$ and $P_{EP}$$>$$P_{B}$. 
                In contrast to the events where magnetic field and density depressions are observed close to the shock, these events
          may result from the { encounter of an intense population of particles accelerated by a traveling shock with} a
       region already characterized by low magnetic field magnitude ($<$5 nT) and low solar wind density ($<$3 cm$^{-3}$) as shown in the examples of Figure 6.

 In contrast to the event not associated with the passage of an interplanetary  shock (2012/025),
                   where the pressure was dominated by $P_{TH}$ and $P_{EP}$ remained small (but $P_{EP}$$>$$P_{B}$ for $\sim$4.8 h; cf. Figure~3),
                    the events associated with the passage of shocks showed continuous exponential increases of $P_{SUM}$
                     in the immediate upstream region of the shocks (indicated by the dashed green lines in Figures 4-7). 
                     Significant are the examples on 2012/029, 2013/206 and 2013/234 where, 
                     regardless of individual changes in $P_{B}$, $P_{TH}$ and $P_{EP}$, $P_{SUM}$ evolved continuously. 
                     Similarly, the event on 2012/025 (without any associated shock) showed a continuous evolution of $P_{SUM}$.  
                     This observation suggests that there is an actual partition among $P_{B}$, $P_{TH}$ and $P_{EP}$
                      that mutually balances their contributions resulting in a smooth evolution of $P_{SUM}$ with time.

The quasi-exponential increase of $P_{SUM}$ observed prior to the passage of the shocks 
        (green dashed lines in Figures 4-7) may result from either the dominant role that $P_{EP}$ has in front of the shocks
        or the existing coupling among $P_{B}$, $P_{TH}$ and $P_{EP}$.
         Whereas in Figure 4 the increase is dominated by the typical exponential increase of particle intensities observed in 
         association with the passage of strong shocks \citep[e.g.,][]{giacalone12},
         in Figure 5 it is observed even when $P_{EP}$ did not show a regular continuous increase. 
If the structure associated with the increase of $P_{SUM}$ 
is simply ``frozen'' into the solar wind,  the quasi-exponential upstream rise of $P_{SUM}$ implies a convected exponential radial gradient 
         $\partial{P_{SUM}}/\partial{r}$$>$0, and that implies that the structure that contains the enhanced energetic particle pressure 
         applies a radially outward force ($-\vec{\nabla}{P_{SUM}}$) on the plasma upstream of the shock. 
         This force can be maintained by the mobile energetic particles streaming upstream of the shock. 
         The energetic ions are quite effective at generating intensity-gradient-driven currents transverse
          to the magnetic field lines, thereby producing the plasma body force $\vec{J}$$\times$$\vec{B}$$\approx$$\vec{\nabla}{P_{EP}}$.  
          This body force would tend to distend the field lines (i.e., it is diamagnetic), thus reducing the magnetic pressure $P_{B}$, 
          and expand (and thereby rarefy) the plasma, thus reducing the thermal pressure $P_{TH}$.

  It is also possible that the process responsible for these depletions of magnetic field and solar wind 
  density in front of interplanetary shocks is similar to some of those proposed for the formation
   of diamagnetic field cavities observed upstream of the Earth's bow shock \citep[e.g.,][and references therein]{sibeck01}. 
   Beams of energetic particles propagating away from the shock may drive magnetic field waves
   that carry magnetic flux to the edges of the beam, resulting in a pair of fast mode compressional and rarefaction
   waves propagating away from the beam axis.
Simulations of this process  result in the creation of a crater of tenuous plasma and weak magnetic field bounded
    by enhanced densities and field strengths \citep{thomas88}.
    We should indicate, however, that the field depressions observed in the Earth's bow shock are quite small
    \citep[$\sim$10-20$\%$ of the ambient field; e.g.,][]{sibeck01}. 
    Here we are reporting, for the most intense events, field depressions of an order of magnitude or more from the ambient field levels.

Hot compressed plasma is observed in the downstream region of the shocks (Figures 4-7).
  The dominance of the shocked thermal plasma behind the shocks indicates that 
     the bulk of the energy given up when the flow is decelerated across the shock goes into thermal plasma, 
             rather than into the energetic particles as, for example, in the case of 
             suprathermal-proton-mediated heliospheric termination shock \citep[e.g.,][]{richardson08}. 
The hot compressed plasma with increased magnetic field in the downstream region of the shocks 
does not allow the creation of diamagnetic cavities by particles propagating in the downstream region of the shocks. 
    Therefore, these diamagnetic cavities are only observed upstream of the shocks when the intensity of energetic particles
     near the shock is large enough to create enough pressure to depress the foreshock magnetic field strength and plasma density. 
  
{    We should indicate that what is measured at the spacecraft is not the gradient of $P_{SUM}$ but the time variation
  of $P_{SUM}$ which is related to both temporal and spatial variations in its proper frame.
  These two terms cannot be distinctly separated with a single spacecraft observation.
  Multipoint measurements  by spacecraft properly distributed in front of a shock would allow us to properly estimate the gradient of $P_{SUM}$
  and relate it to other forces acting in the upstream medium of the shock.
  For example, \citet{oka11} used the close alignment of five proves in the Earth's magnetotail
  to show that there is a radially outward pressure gradient  existing both in stationary and perturbed conditions (see their Figure 4).
  In stationary condition, the pressure gradient force is balanced by an earthward tension force,
  whereas in disturbed conditions,  the X-line points forming in the magnetotail move downtail away from Earth.
  In the case of propagating interplanetary shocks that continuously accelerate particles, it is difficult to determine if
   there is a force that can balance the radially outward force $-\vec{\nabla}{P_{SUM}}$. 
 As the shock propagates away from the Sun and its efficiency in accelerating particles decreases, 
 the pressure gradient resulting from particle streaming would weaken, 
 and it is only in the upstream region of fast strong shocks accompanied by intense energetic particle populations where the
 effects of $P_{EP}$ can be observed.    
  }      

It has been argued that the pressure exerted by energetic particles ahead of
a shock can also decelerate the background medium where the shock propagates
  \citep[e.g.,][and references therein]{axford77, drury81, eichler81, terasawa06}. 
            The most intense events in Figure 1 clearly show a steady increase in the solar wind
             speed as the shock approached the spacecraft (i.e. a decrease of the flow speed in the frame comoving with the shock).
The event in Figure~5 did not show a decrease in the flow speed just before the shock,
whereas the event in Figure~4 shows a significant increase of $V_{sw}$ as $P_{EP}$ increases 
              Similarly, the events in Figures 6 and 7 did not show a clear pattern in the evolution of $V_{sw}$. 
              Therefore, we conclude that the decrease of the flow speed is only observed in the most intense events of our sample. 

 The event on 20 October 1989 showed a weakened shock with a density compression ratio of, at most, $r_{n}$$\sim$2 \citep{lario02} 
              because of the gradual increase of $n_{p}$ observed $\sim$30 minutes before the shock arrival. 
              The increase of $n_{p}$ and $B$ just a few minutes before the arrival of shocks is also observed in less intense events 
              (e.g., for $\sim$5 minutes in the shock on 2012/149 and for $\sim$30 minutes in the shock on 2013/206), 
              but the compression ratios of these two shocks were still relatively high (Table~1).
              Therefore, it seems that any possible weakening of the shock resulting from this
               precursor increase of $n_{p}$ and $B$ can only be observed in the most intense events.

To summarize, we would like to emphasize that
the frequency of events where $P_{EP}$ is larger than, or comparable to, $P_{B}$ and $P_{TH}$ is quite significant (see Figures 2e and 2f).
The dynamic effects that energetic particles produce on the medium upstream of fast interplanetary shocks
may lead to modifications of the shock properties relative to those of a shock propagating into a medium without energetic particle effects.
In general, the contribution made by the energetic particles to the total pressure measured
                   in the upstream region of  shocks is not negligible, and in many cases may exceed that of the magnetic field and the thermal plasma.
                          Computations of shock parameters usually neglect the effect that $P_{EP}$ may have on the shock properties.
                           Given the frequency of events with upstream regions dominated by $P_{EP}$,
                            or where the contribution of $P_{EP}$ to the total pressure is significant,
                             the correct computation of shock parameters may require the use of generalized Rankine-Hugoniot
                              conservation equations across the shocks that include the contributions of the
                               non-thermal energetic particles in the conservation of momentum and energy across the shocks \citep[e.g.,][]{roelof10}.

\acknowledgments We acknowledge the STEREO Science Centers for providing the data used in this paper and the
 STEREO/IMPACT and STEREO/PLASTIC teams for providing these data. 
 DL acknowledges the support from NASA under grants NNX11A083G and NNX15AD03G. 
 ECR and DL acknowledge NASA support under the ACE grant NNX10AT75G. 
 DL acknowledges the International Space Science Institute (ISSI) at Bern, Switzerland,
  for their funding of the team ``Exploration of the inner Heliosphere: what we have learned
   from Helios and what we want to study with Solar Orbiter," led by Dr. W. Droege, and the useful discussions with the team members.

\clearpage



\begin{figure}
\epsscale{.60}
\plotone{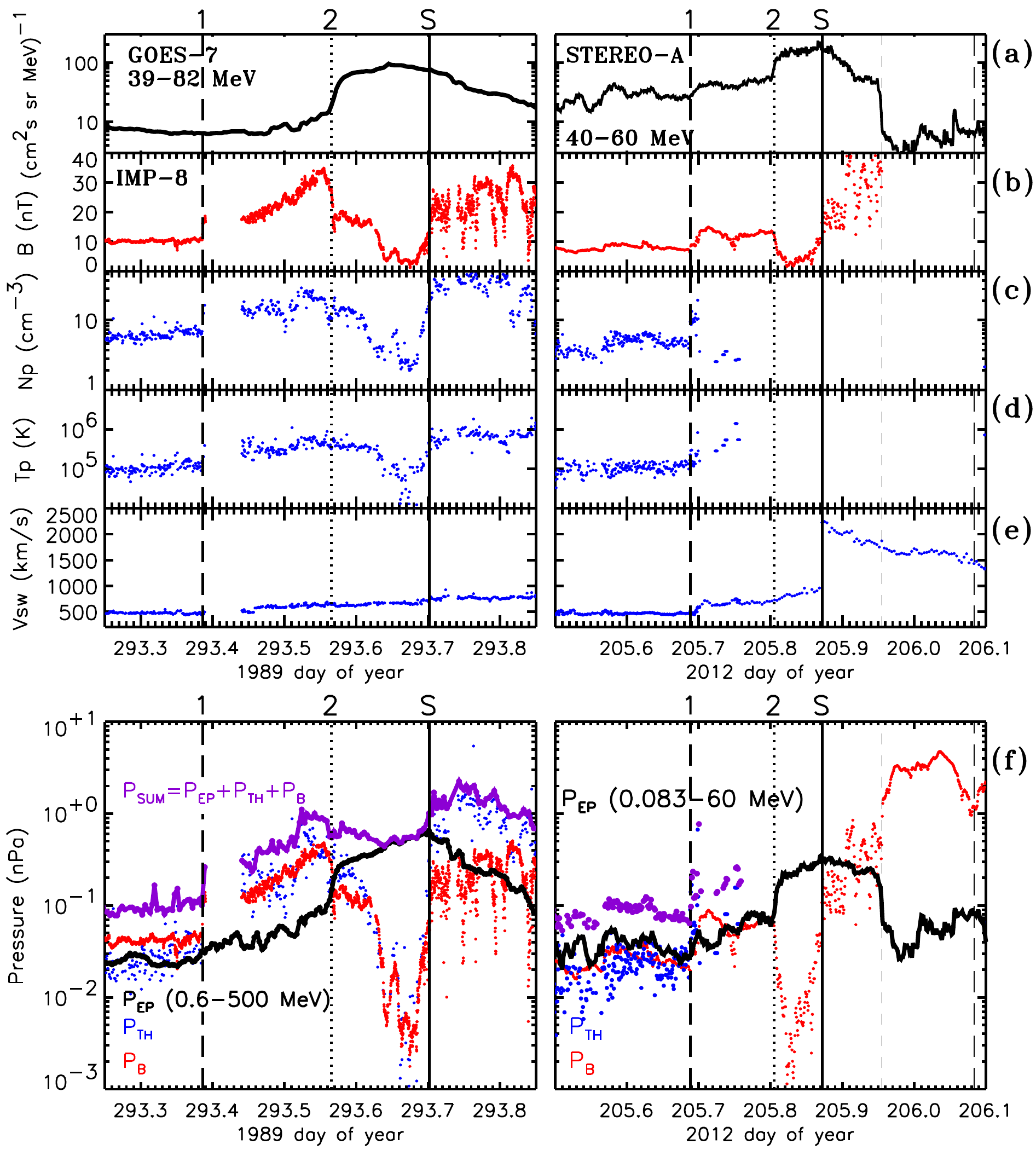}
\caption{(Left) GOES-7 and IMP-8 observations during the event associated
 with the passage of the interplanetary shock on 20 October 1989 (day of year 293). 
 (Right) STEREO-A observations during the event associated with the passage 
 of the interplanetary shock on 23 July 2012 (day of year 205). 
 (a) 5-minute averages of 39-82 MeV proton intensities measured by GOES-7
  and 1-minute averages of 40-60 MeV proton intensities measured by STEREO-A/IMPACT/HET, 
  (b) magnetic field magnitude, (c) solar wind proton density, (d) solar wind proton temperature, 
  (e) solar wind proton speed, (f) magnetic field pressure (red), solar wind thermal pressure (blue), 
  energetic particle partial pressure (black) computed over the indicated energy range, 
  and the sum $P_{EP}$+$P_{B}$+$P_{TH}$ (purple trace). The solid vertical lines labeled with the symbol S indicate the passage of the shocks.
   The dashed vertical lines labeled with the number 1 indicate the passage of a prior interplanetary shock as identified by \citet{lario02} and \citet{russell13}. 
   The dotted vertical lines labeled with the number 2 indicate the increase of energetic particles before the shock passage
    that was accompanied by a depletion of the magnetic field. 
    The time interval between the dotted line and the shock S is when $P_{EP}$ exceeded $P_{B}$ by more than one order of magnitude.\label{fig1}}
\end{figure}

\clearpage

\begin{figure}
\epsscale{0.8}
\plotone{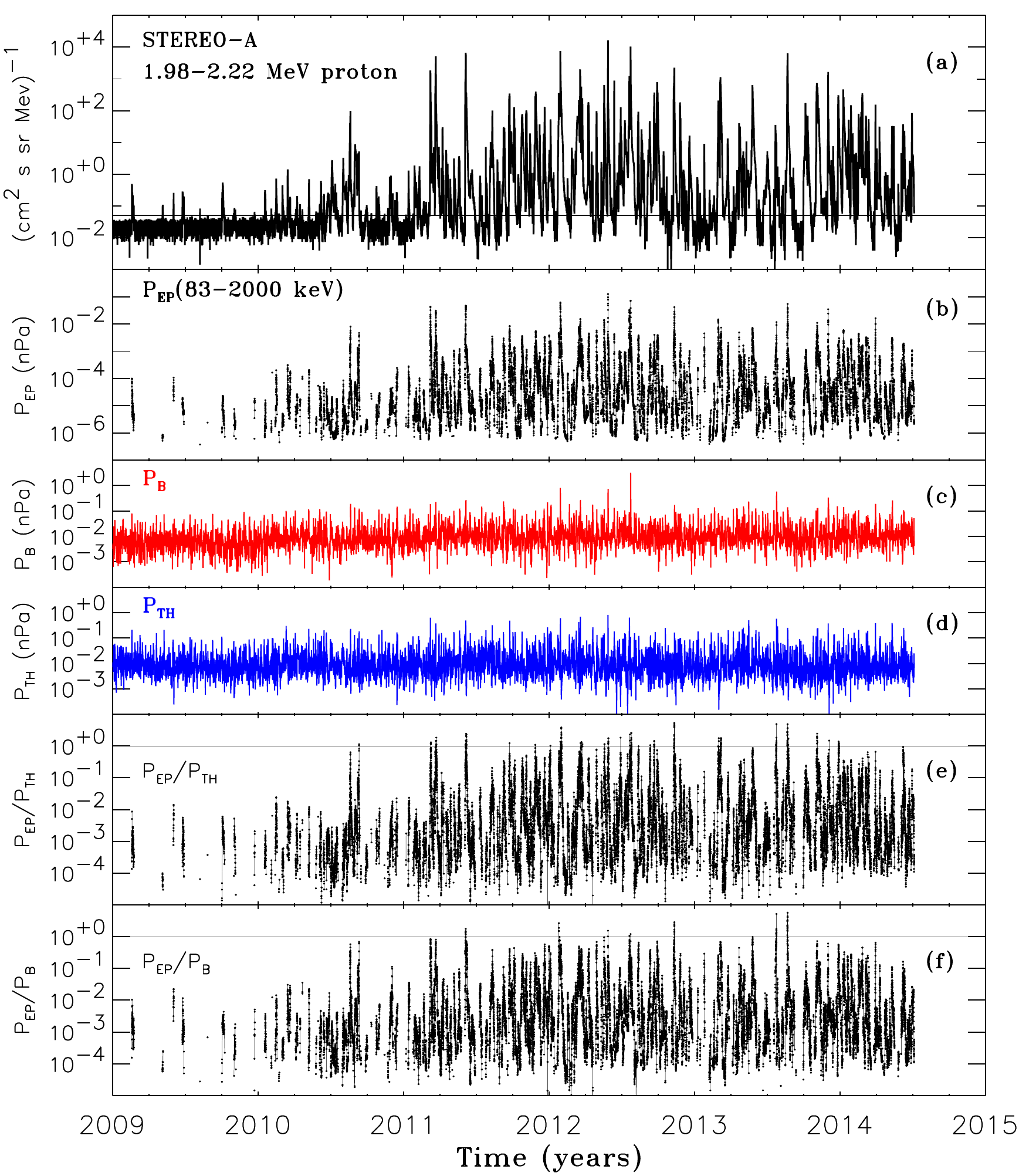}
\caption{Hourly averages of (a) 1.98-2.22 MeV proton intensities measured by STEREO-A/IMPACT/SEPT, 
(b) the pressure $P_{EP}$ exerted by protons in the energy range 83-2000 keV, (c) magnetic field pressure $P_{B}$, 
(d) solar wind thermal pressure $P_{TH}$, (e) the ratio between $P_{EP}$ and $P_{TH}$, and (f) the ratio between $P_{EP}$ and $P_{B}$. 
The horizontal straight line in (a) indicates the differential intensity value above which $P_{EP}$ has been computed. 
The horizontal straight lines in (e) and (f) indicate the values $P_{EP}$/$P_{TH}$ = 1 and $P_{EP}$/$P_{B}$ = 1, respectively.\label{fig2}}
\end{figure}

\begin{figure}
\epsscale{.70}
\plotone{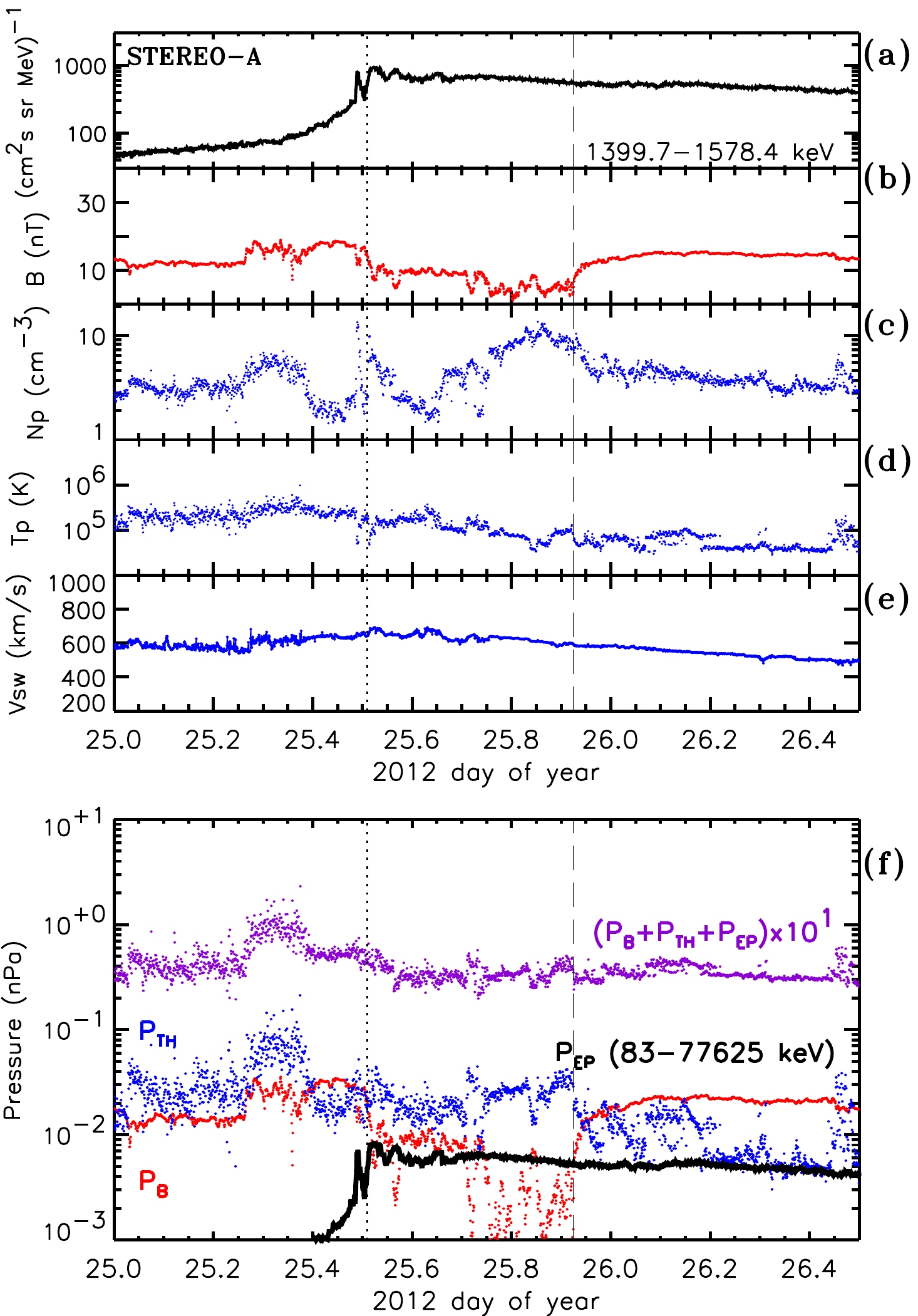}
\caption{STEREO-A observations during the event on 2012/025. 
One-minute averages of (a) 1.4-1.6 MeV proton intensities, (b) magnetic field magnitude, 
(c) solar wind density, (d) solar wind proton temperature, (e) solar wind speed, (f) magnetic field pressure $P_{B}$ (red), 
solar wind thermal pressure $P_{TH}$ (blue), energetic particle partial pressure $P_{EP}$ (black)
 computed over the indicated energy range, and the sum $P_{EP}$+$P_{B}$+$P_{TH}$ (purple trace). 
 The dashed vertical line indicates the entry of STEREO-A into an ICME. 
 The dotted vertical line indicates the onset of the time interval with depressed magnetic
  field strength and elevated energetic particle intensities. \label{fig3}}
\end{figure}

\begin{figure}
\epsscale{.65}
\plotone{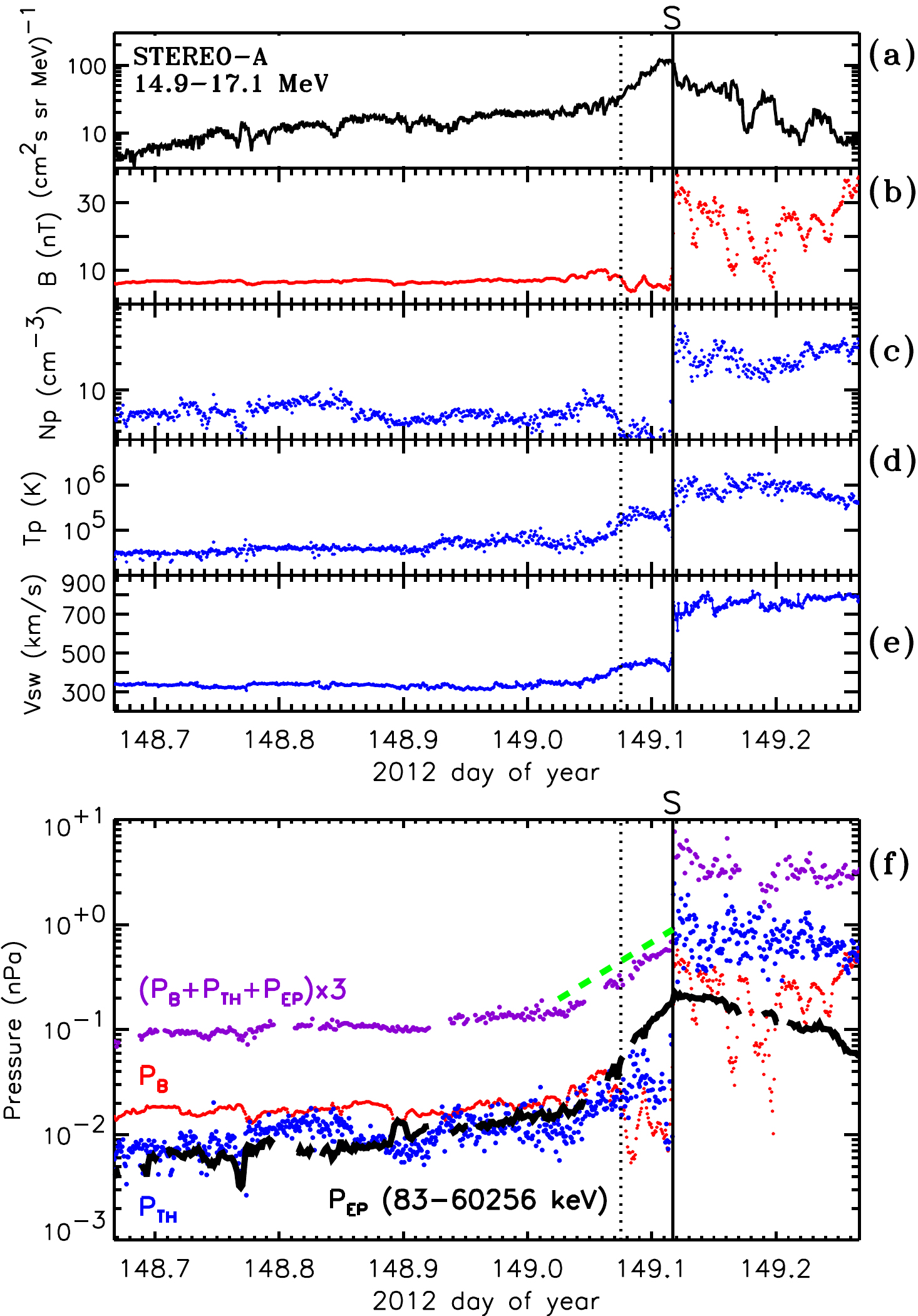}
\caption{STEREO-A observations during the event associated with the arrival
 of an interplanetary shock on 2012/149. One-minute averages of
  (a) 14.9-17.1 MeV proton intensities, (b) magnetic field magnitude, 
  (c) solar wind density, (d) solar wind proton temperature, (e) solar wind speed, 
  (f) magnetic field pressure $P_{B}$ (red), solar wind thermal pressure $P_{TH}$ (blue), 
  energetic particle partial pressure $P_{EP}$ (black) computed over the indicated energy range, 
  and the sum $P_{EP}$+$P_{B}$+$P_{TH}$ (purple trace). The solid vertical line indicates the passage of the interplanetary shock. 
  The dotted vertical line indicates the onset of the time interval with depressed
   magnetic field, tenuous solar wind density and $P_{EP}$$>$$P_{B}$. 
   The dashed green line in panel (f) indicates the exponential increase of $P_{SUM}$ observed before the arrival of the shock. \label{fig4}}
\end{figure}

\begin{figure}
\epsscale{.65}
\plotone{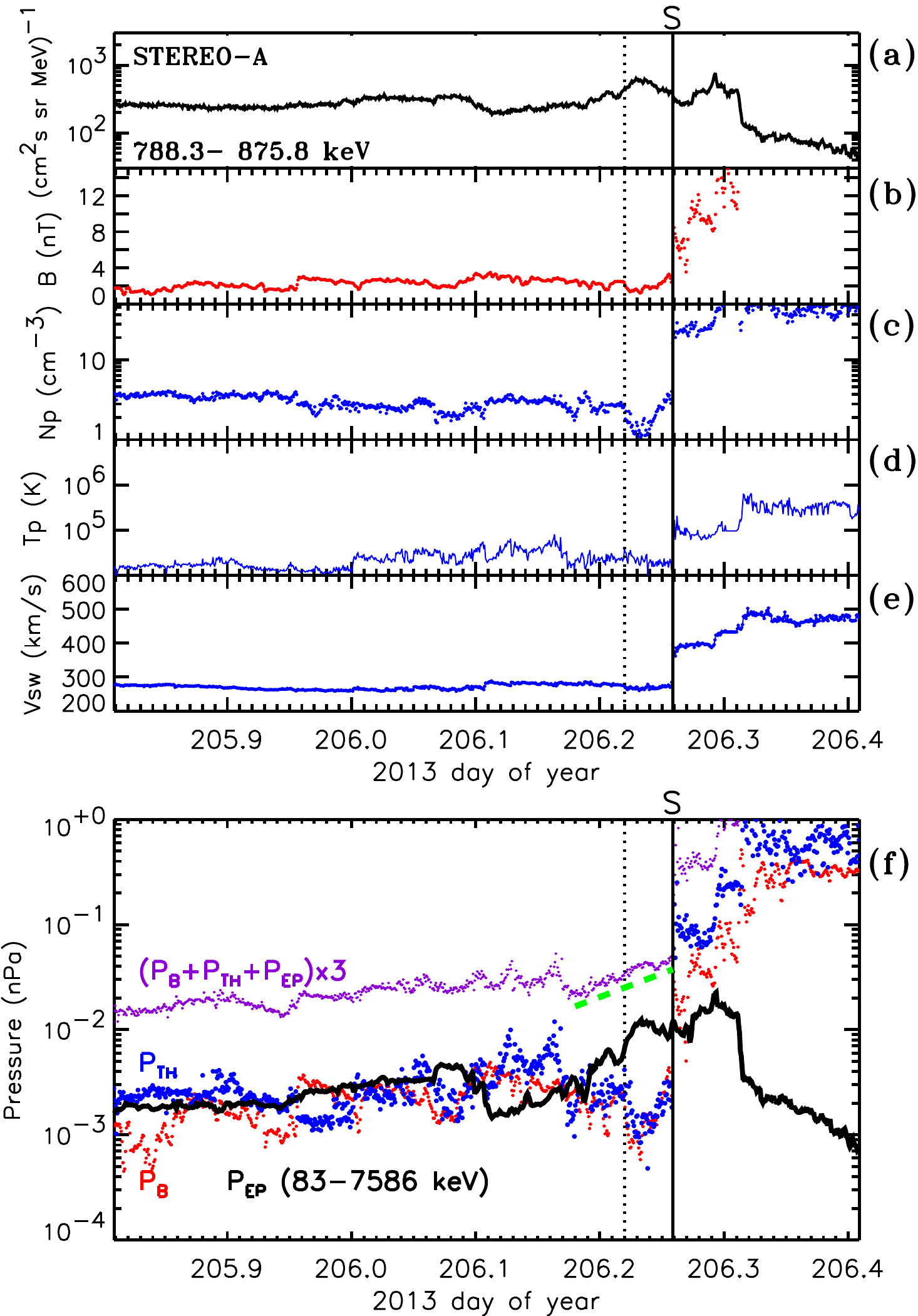}
\caption{The same as Figure 4 but for the event associated with the passage of a shock on 2013/206. \label{fig5}}
\end{figure}

\begin{figure}
\epsscale{.8}
\plotone{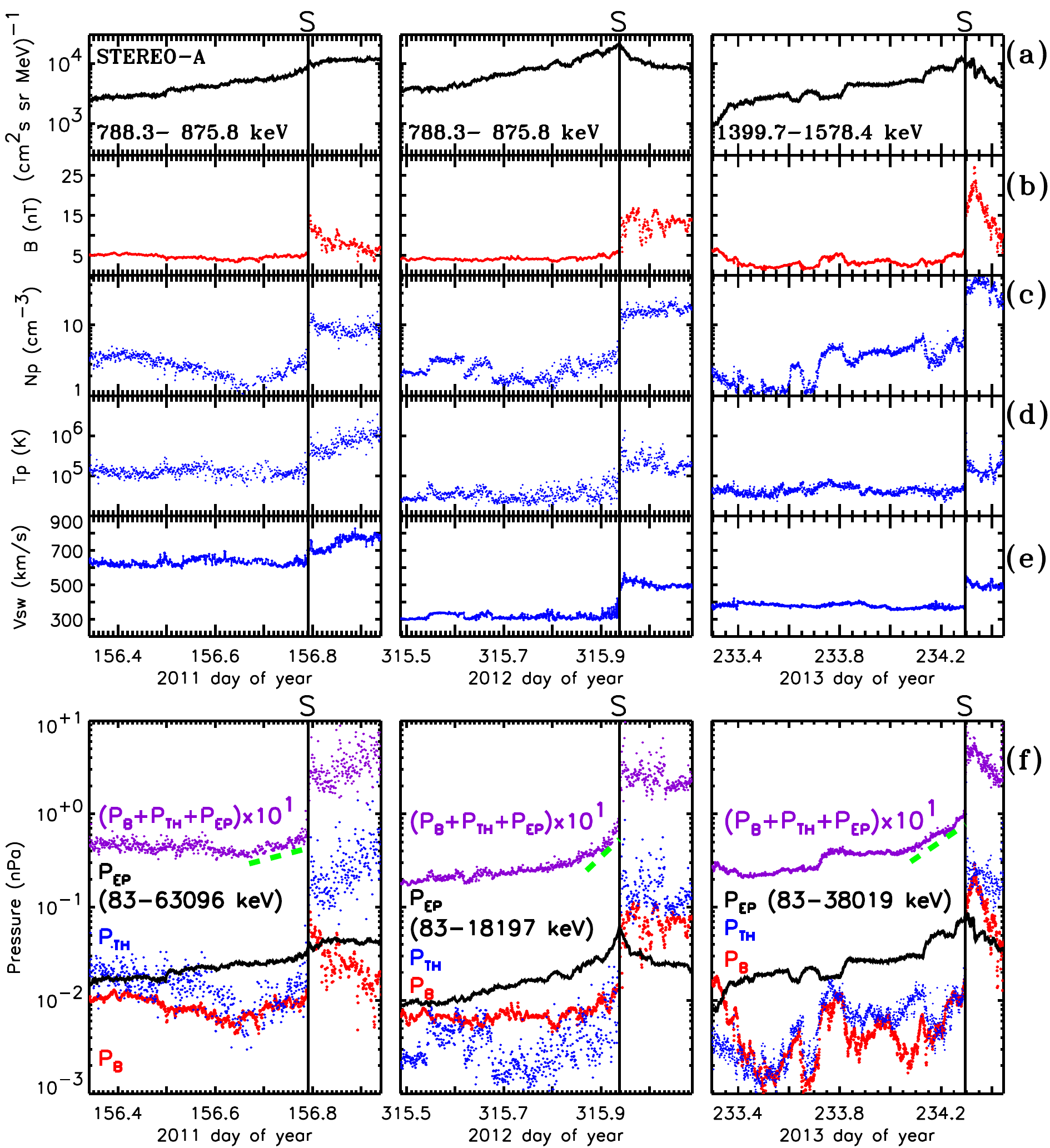}
\caption{STEREO-A observations during the events associated with the passage of the shocks on 2011/156, 2012/315 and 2013/234, 
depicted in the same format as in Figure 4.\label{fig6}}
\end{figure}

\begin{figure}
\epsscale{.8}
\plotone{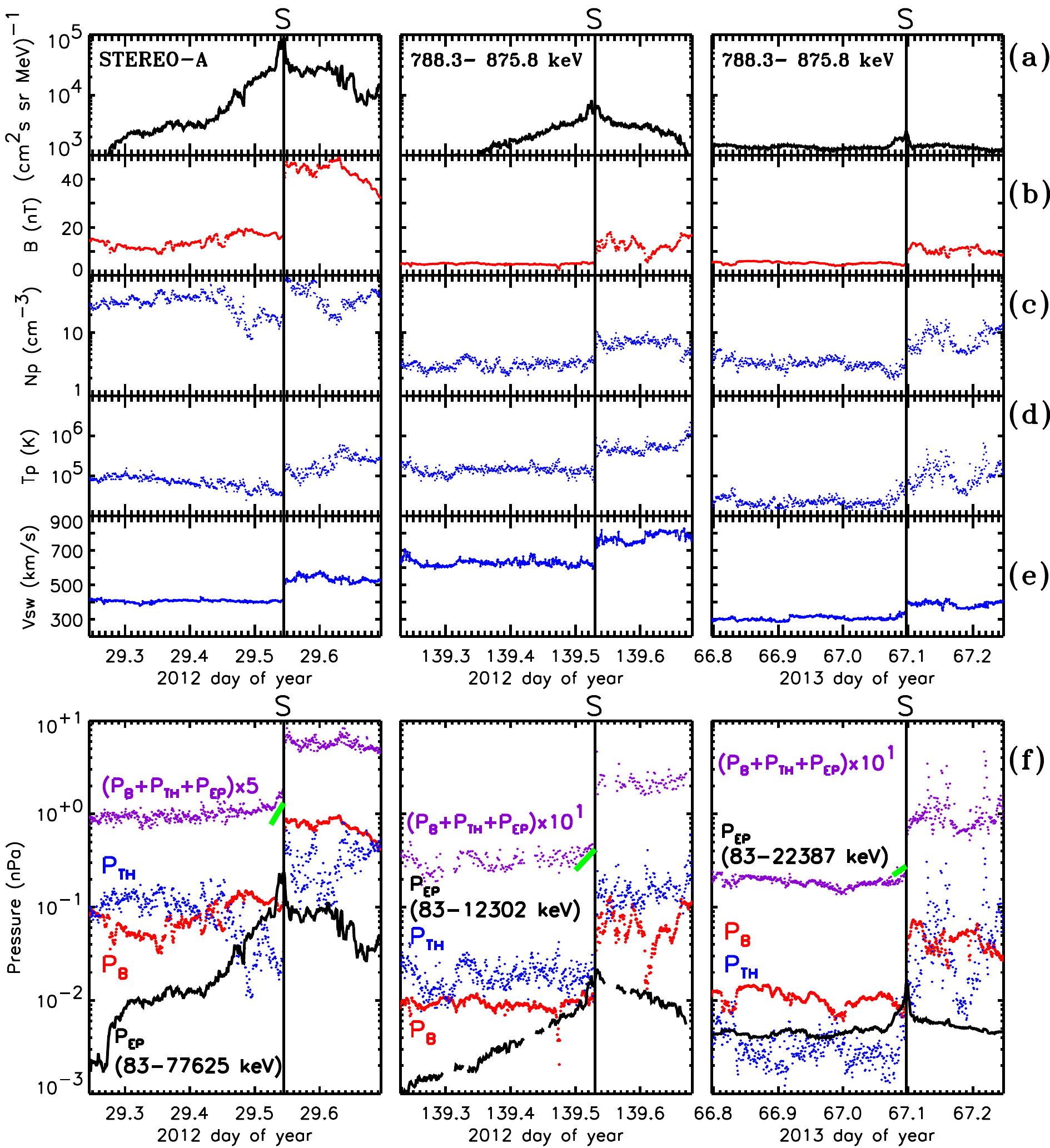}
\caption{STEREO-A observations during the events associated with the passage of the shocks on 2012/029, 2012/139 and 2013/067, 
depicted in the same format as in Figure 4. \label{fig7}}
\end{figure}

\clearpage

\begin{table}
\scriptsize{
\begin{center}
\caption{Time intervals when $P_{EP}$$>$$P_{B}$ obtained using hourly averages 
of magnetic field and energetic particle measurements from the magnetometer and the SEPT particle detector of the STEREO-A/IMPACT instrument suite.\label{tbl-1}}
\begin{tabular}{lccll}
\tableline\tableline
Date & Duration & Association & Shock compression  & \multicolumn{1}{c}{$\tau$}\\
(Year/Day) & (min) &  & ratios $r_{B}$ and $r_{n}$ $^{a}$ & (day)$^{c}$\\
 \tableline
2011/156	& 750	& shock	& $r_{B}$=3.07 $r_{n}$=3.82  & 0.318 \\
2012/025	& 288	& depressed $B$	&  & \\
2012/149	& 60	& shock	& $r_{B}$=6.81 $r_{n}$=9.01& 0.060  \\
2012/205	& 96	& shock	& $r_{B}$=2.17$^{b}$    &  \\
2012/315	& 882	& shock	& $r_{B}$=2.06 $r_{n}$=5.19 & 0.088 \\
2013/206	& 100	& shock	& $r_{B}$=4.19 $r_{n}$=5.19 & 0.097 \\
2013/234	& 1373	& shock	& $r_{B}$=3.18 $r_{n}$=3.39 & 0.235 \\
\tableline
\end{tabular}
\tablenotetext{a}{Compression ratios obtained using the \citet{vinas86} technique.}
\tablenotetext{b}{Magnetic compression ratio provided by J. Lian (www-ssc.igpp.ucla.edu/$\sim$jlian/STEREO/Level3)}
\tablenotetext{c}{Time scale of the exponential increase of $P_{SUM}$ obtained by fitting a function $P_{SUM}(t)=P_{SUM}(t_{1})\exp{\Bigl\{(t-t_{1})/\tau\Bigr\}}$
before the arrival of the shocks.}
\end{center}
}
\end{table}


\clearpage






\end{document}